\documentclass[aps,prl,twocolumn,showpacs,showfootnotes]{revtex4}
\usepackage{graphicx}
\usepackage{latexsym}
\def\beq{\begin{equation}}
\def\eeq{\end{equation}}
\def\bey{\begin{eqnarray}}
\def\eey{\end{eqnarray}}

\def\lsim{\mathrel{\raise.3ex\hbox{$<$\kern-.75em\lower1ex\hbox{$\sim$}}}}
\def\gsim{\mathrel{\raise.3ex\hbox{$>$\kern-.75em\lower1ex\hbox{$\sim$}}}}

\newcommand{\be}{\begin{equation}}
\newcommand{\ee}{\end{equation}}

\begin{document}

\title{The PAMELA and ATIC Excesses From a Nearby Clump of Neutralino Dark Matter}  
\author{Dan Hooper$^{1,2}$, Albert Stebbins$^{1}$, and Kathryn M. Zurek$^1$}
\address{$^1$Theoretical Astrophysics, Fermi National Accelerator Laboratory, Batavia, IL \\ $^2$Department of Astronomy and Astrophysics, The University of Chicago, Chicago, IL}

\date{\today}

\begin{abstract}

In this letter, we suggest that a nearby clump of 600-1000 GeV neutralinos may be responsible for the excesses recently observed in the cosmic ray positron and electron spectra by the PAMELA and ATIC experiments. Although neutralino dark matter annihilating throughout the halo of the Milky Way is predicted to produce a softer spectrum than is observed, and violate constraints from cosmic ray antiproton measurements, a large nearby (within 1-2 kiloparsecs of the Solar System) clump of annihilating neutralinos can lead to a spectrum which is consistent with PAMELA and ATIC, while also producing an acceptable antiproton flux. Furthermore, the presence of a large dark matter clump can potentially accommodate the very large annihilation rate required to produce the PAMELA and ATIC signals.

\end{abstract}
\pacs{95.35.+d; 95.85.Ry; FERMILAB-PUB-08-566-A}
\maketitle

Recent observations by the PAMELA~\cite{pamela} and ATIC~\cite{atic} experiments have revealed a surprisingly large flux of high energy electrons and positrons in the cosmic ray spectrum. These observations strongly indicate the presence of a relatively local source of energetic pairs. Although the origin of these particles remains unknown, a nearby pulsar~\cite{pulsars,pulsars2} and dark matter annihilations~\cite{dm,ourpaper,lepsom,leptons,antiprotons,antiprotons2,gammarayssyn,sommerfeld} have each been proposed as possible sources. 

Efforts to explain the PAMELA and ATIC excesses with annihilating dark matter have faced a number of challenges, however. In particular:
\begin{itemize}
\item{The spectrum of electrons and positrons predicted to be generated in the annihilations of most dark matter candidates is much too soft to fit the observations of PAMELA and ATIC. If WIMPs annihilating throughout the halo of the Milky Way are to produce the spectral shape observed by these experiments, they must annihilate mostly to charged leptons ($e^+ e^-$, $\mu^+ \mu^-$, and/or $\tau^+ \tau^-$)~\cite{ourpaper}. While models have been proposed in which this is the case~\cite{lepsom,leptons}, many of the most often studied WIMP candidates (including neutralinos) are predicted to annihilate dominantly to quarks and/or gauge bosons~\cite{neu}.}
\item{The dark matter annihilation rate that is required to generate the observed spectrum of cosmic ray electrons and positrons is much higher (by a factor of $\sim$$10^2$-$10^3$) than is predicted for a typical thermal relic distributed smoothly throughout the Galactic Halo~\cite{ourpaper}. To normalize the annihilation rate to the PAMELA and ATIC signals, we must require either large inhomogeneities in the dark matter distribution which lead to a considerably enhanced annihilation rate ({\it ie.} a `` boost factor''), and/or dark matter particles which possess a considerably larger annihilation cross section than is required of a thermal relic. This latter option requires either a non-thermal production mechanism in the early universe, or an enhancement of the annihilation cross section at low velocities, such as through the Sommerfeld effect~\cite{lepsom,sommerfeld}.}
\item{The very large annihilation rate required to generate the PAMELA and ATIC signals can also lead to the overproduction of cosmic ray antiprotons~\cite{antiprotons,antiprotons2}, gamma rays, and synchrotron emission~\cite{gammarayssyn}. For example, Ref.~\cite{antiprotons2} finds that a 1 TeV WIMP annihilating to $W^+ W^-$ would be expected to exceed the observed cosmic ray antiproton flux by a factor of approximately 5-10 if the overall annihilation rate is normalized to the PAMELA positron fraction.}
\end{itemize}

Collectively, these considerations appear to strongly limit the type of dark matter particle that could potentially generate the observed spectra of cosmic ray electrons and positrons. Under the astrophysical assumptions that are typically adopted, dark matter must annihilate almost entirely to charged leptons, and at a very high rate, if they are to accommodate the observations of PAMELA and ATIC. 

Here, we propose an alternative scenario that can alleviate all three of the challenges listed above.  In particular, we consider the case in which the high energy cosmic ray electron and positron spectra are dominated by neutralino annihilations taking place in a nearby clump of dark matter. In such a scenario, WIMPs which annihilate to non-leptonic final states can still generate the spectra observed by PAMELA and ATIC, without violating constraints from cosmic ray antiprotons, gamma rays, or synchrotron emission. 

\begin{figure}
\resizebox{9.0cm}{!}{\includegraphics{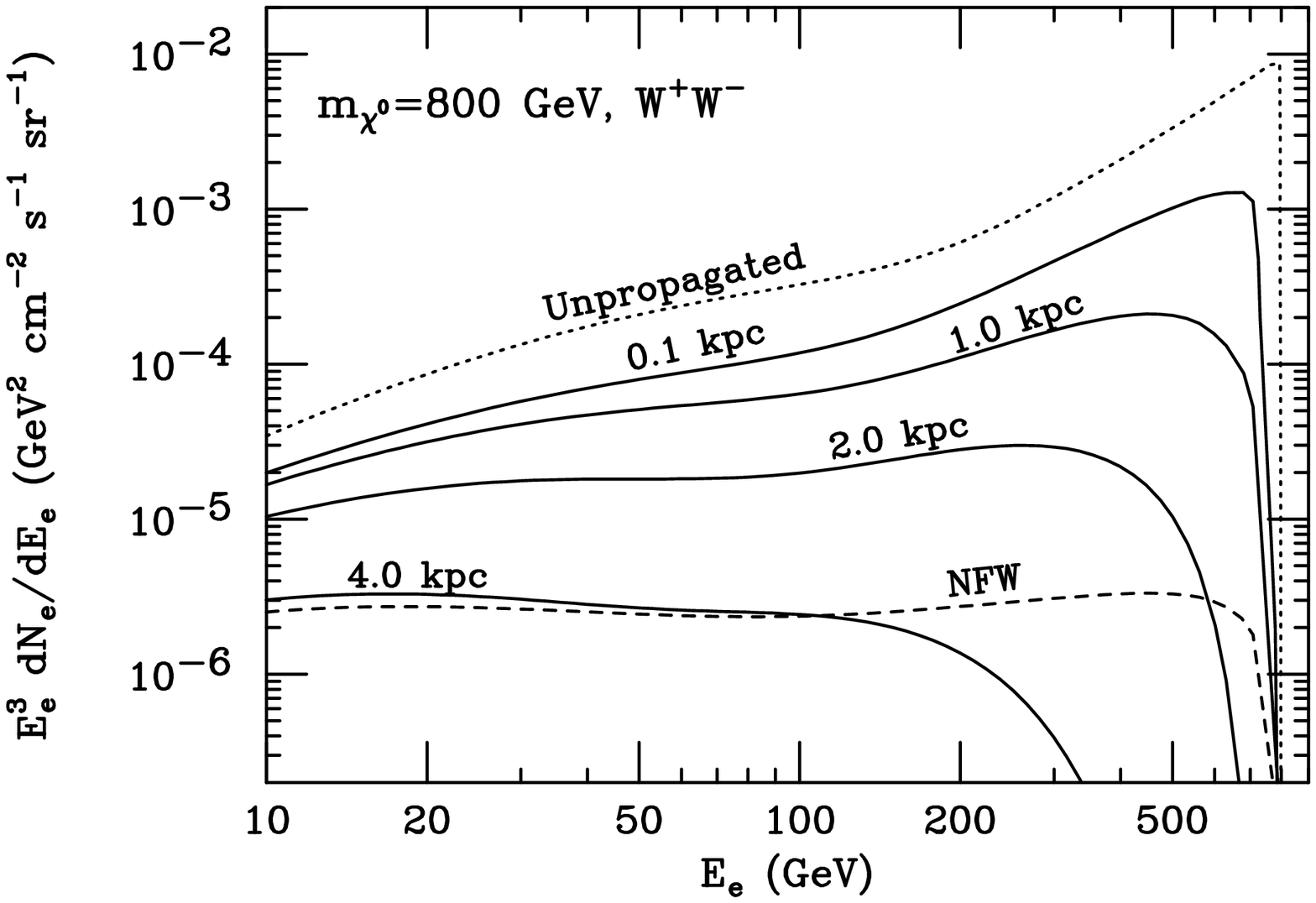}} \\
\resizebox{9.0cm}{!}{\includegraphics{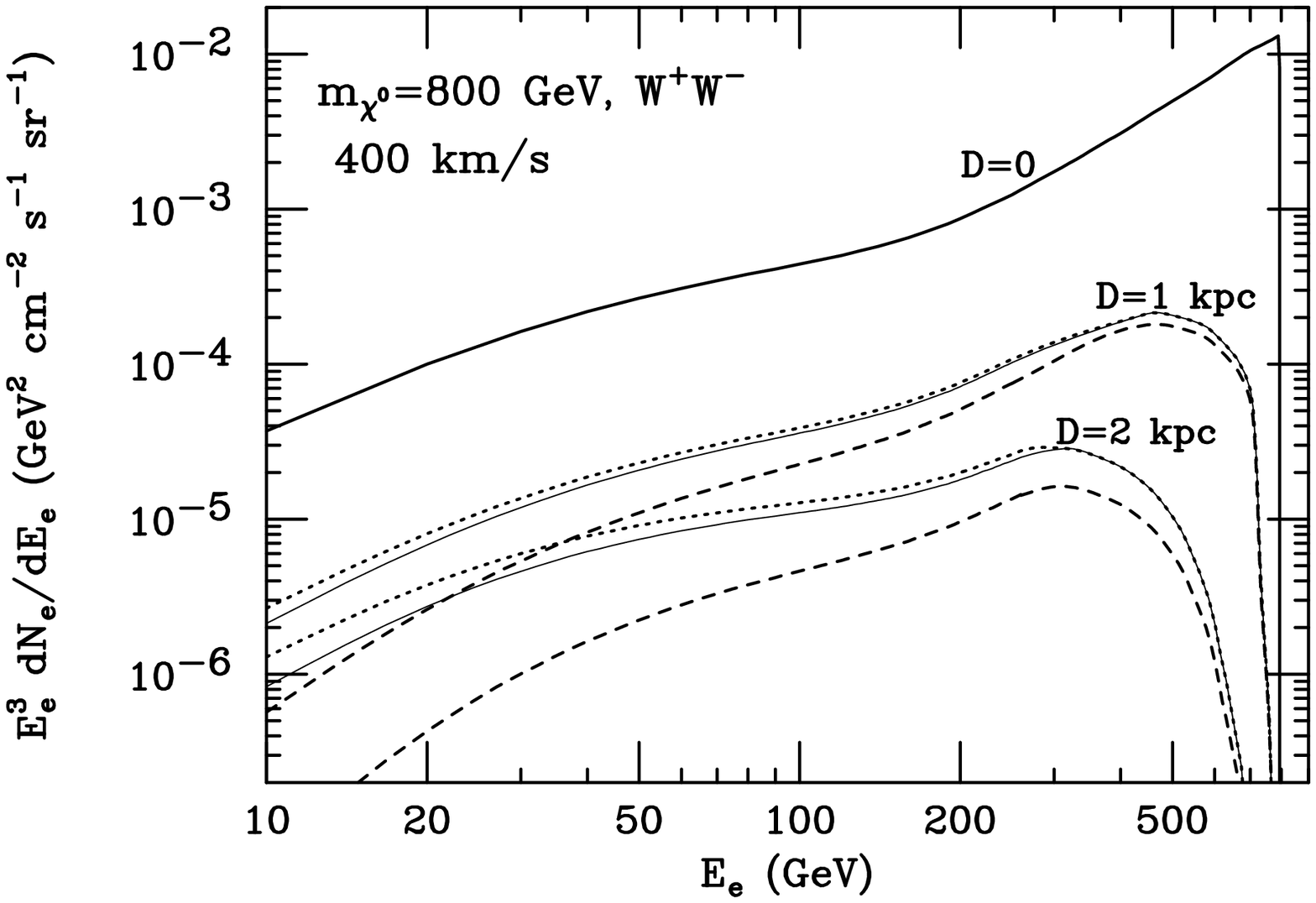}}
\caption{The electron plus positron spectrum from a clump of 800 GeV neutralinos annihilating to $W^+W^-$. In the top frame, the clump is assumed to be stationary and at distances of 0.1, 1, 2 or 4 kpc from the Solar System. For comparison, we show with arbitrary normalization the spectral shape prior to propagation (dotted line) and the result for a smooth NFW halo profile (dashed line). In the lower frame, results are shown for a clump moving at 400 km/s relative to the Solar System. The thick solid line denote the case in which the clump has recently reached the Solar System. Other line types describe the case in which a clump which has passed through the Solar System and is now 1 or 2 kpc away (dotted, top-to-bottom), a clump which is approaching the Solar System and is currently 1 or 2 kpc away (dashed, top-to-bottom), and a clump which passed with a closest approach of 1 kpc away from the Solar System and is now $\sqrt{2}$ or $2\sqrt{2}$ kpc away (thin solid). In each case shown, an annihilation rate of $2 \times 10^{35}$ s$^{-1}$  was used.}
\label{fig1}
%\end{center}
\end{figure}

As electrons and positrons propagate through the radiation fields and magnetic field of the Milky Way, they lose energy through inverse Compton and synchrotron processes. As a result, the spectrum from distant objects is softened relative to that originating from more local sources.  This is illustrated in the top frame of Fig.~\ref{fig1}, where we show the spectrum of electrons plus positrons from the annihilations of WIMPs in a stationary clump 0.1, 1, 2, or 4 kpc from the Solar System. Here, we have considered an 800 GeV WIMP which annihilates to $W^+ W^-$ (such as a wino-like neutralino, for example). We compare this to the spectral shape predicted from annihilations throughout a smooth (Navarro-Frenk-White, NFW~\cite{nfw}) halo profile, and to the spectral shape of electrons/positrons prior to propagation.  From this, it is clear that a nearby clump of annihilating dark matter can lead to a spectrum that is considerably harder than is predicted from annihilations throughout the halo at large.

To calculate the cosmic ray electron/positron spectrum taking into account the effects of diffusion and energy losses, we have solved the propagation equation~\cite{prop}:
\begin{eqnarray}
\frac{\partial}{\partial t}\frac{dN_{e}}{dE_{e}} &=& \vec{\bigtriangledown} \cdot \bigg[K(E_{e},\vec{x})  \vec{\bigtriangledown} \frac{dN_{e}}{dE_{e}} \bigg]
+ \frac{\partial}{\partial E_{e}} \bigg[b(E_{e},\vec{x})\frac{dN_{e}}{dE_{e}}  \bigg] \nonumber \\ &+& Q(E_{e},\vec{x}),
\label{dif}
\end{eqnarray}
where $dN_{e}/dE_{e}$ is the number density of positrons per unit energy, $K(E_{e},\vec{x})$ is the diffusion coefficient, and $b(E_{e},\vec{x})$ is the rate of energy loss. The source term, $Q(E_e, \vec{x})$, reflects both the distribution of dark matter in the Galaxy, and the mass, annihilation cross section, and dominant annihilation channels of the neutralino. Throughout this letter, we adopt the following diffusion parameters:   $K(E_{\rm e}) =5.3\times10^{28}\, (E_{\rm e} / 4 \, \rm{GeV})^{0.43} \,{\rm cm^2 /s}$, and $b(E_e) = 10^{-16} \, ({E_e} / 1 \, \rm{GeV})^2 \,\, \rm{s}^{-1}$. We also select boundary conditions corresponding to a slab of half-thickness 4 kpc, beyond which cosmic ray electrons/positrons are allowed to freely escape the Galactic Magnetic Field. 

%It should be noted that constraints on these diffusion parameters are generally made using information from cosmic ray nuclei which carry information... As this information pertains to the diffusion of cosmic rays over much larger volumes than the distance to the clump in our scenario, considerable variations from these quantities may be possible. That being said, these remain reasonable estimates for the calculation at hand.

Dark matter substructures close to the center of a dark matter halo, such as near the Solar System in the Milky Way, typically move with velocities of a few times the rotational velocity ($v_{\rm rot}=220\,${\rm km/s} for the Sun).  We can expect a broad distribution of clump velocities, but typical values are approximately $400\,$km/s.  The stationary solution to Eq.~(\ref{dif}), such as used in Refs.~\cite{taylor,cumberbatch}, is a good approximation when $v R\ll K[E_{\rm e}]$, where $R$ is the distance to the clump and $v$ is the velocity of the clump with respect to the ISM.   Given our choice, $K(E_{\rm e}) \approx 685\,(E_{\rm e} / 100\, \rm{GeV})^{0.43} \,{\rm kpc\,km/s}$,  we see that clump motions are largely unimportant for $(E_{\rm e}/100\,$GeV)$^{0.43}\gg R/$kpc, but increasingly important at lower energies.  In Fig.~1 we show the electron plus positron distribution from dark matter annihilations for a stationary clump and a moving one.  Including the motion of the clump hardens the spectrum further. At energies below a few hundred GeV, the spectrum is suppressed relative to the steady state case. In contrast, the spectrum is unchanged above a few hundred GeV.
%AJS I THINK THIS IS WRONG INTERPRETATION
%, as in this regime the energy loss time is shorter than the duration over which the clump is in the vicinity the Solar System.

\begin{figure}
\resizebox{9.0cm}{!}{\includegraphics{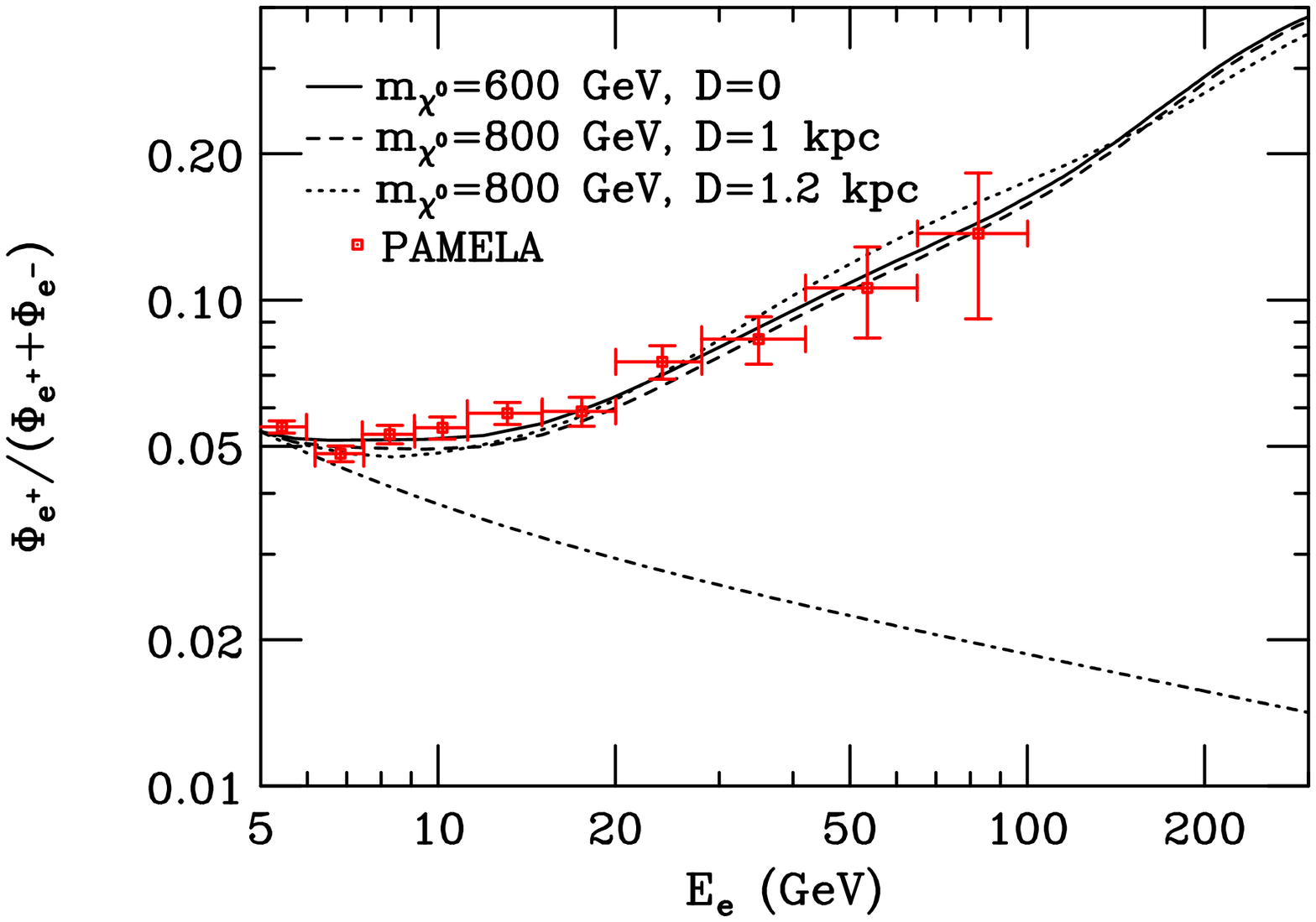}} \\
\resizebox{9.0cm}{!}{\includegraphics{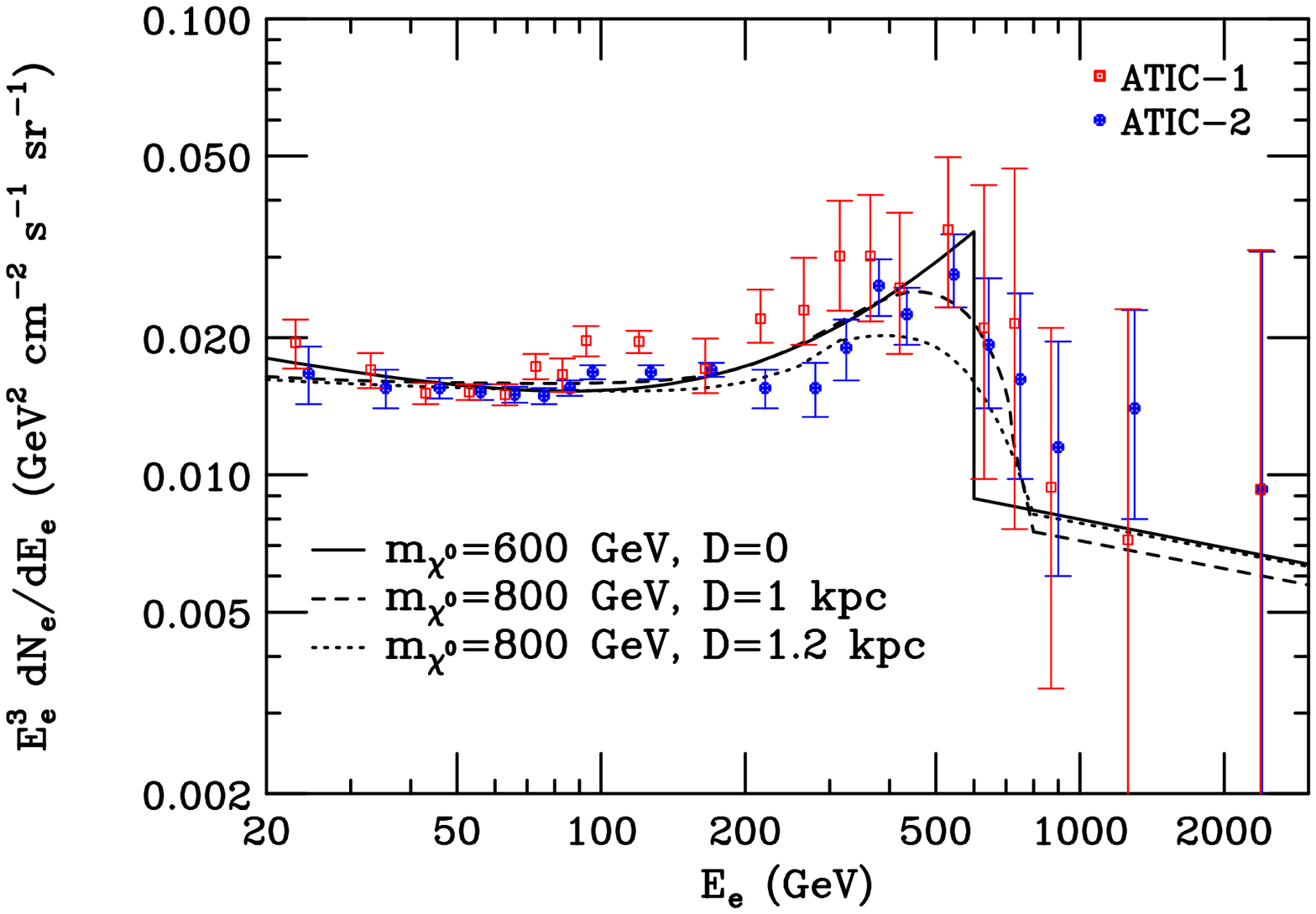}}
\caption{The positron fraction (top) and the electron plus positron spectrum (bottom) from a nearby clump of annihilating neutralino dark matter. The solid lines denote the result for a very nearby clump of 600 GeV neutralinos, while the dashed and dotted lines correspond to 800 GeV neutralinos in a clump 1 and 1.2 kpc away, respectively. To normalize the curves, we used approximate annihilation rates of $7 \times 10^{35}$, $1.7 \times 10^{37}$, and $2.2 \times 10^{37}$ per second for the $D=0,1$, and $1.2$ cases, respectively. Each case provides a good fit to both the PAMELA and ATIC data. For comparison, we show in the top frame as a dot-dashed line the astrophysical expectation from cosmic ray secondary production~\cite{sec}.}
\label{fig3}
%\end{center}
\end{figure}

In Fig.~\ref{fig3}, we compare our results to the measurements of PAMELA (top) and ATIC (bottom). Again, we have considered a dark matter particle annihilating to $W^+W^-$, such as a wino-like neutralino. From these frames, it is clear that a nearby clump of such particles can accommodate the measurements reported by these experiments. As the ATIC feature becomes less prominent for more distant clumps, the origin of the signal must lie within approximately 1 or 2 kpc of the Solar System to generate the observed spectral shape.

To normalize the signals shown in Fig.~\ref{fig3}, we have used annihilation rates of $7\times 10^{35}$, $1.7\times 10^{37}$, and $2.2\times 10^{37}$ per second for clumps at distances of 0, 1 and 1.2 kpc, respectively. These very large annihilation rates require us to consider a very large or dense clump of dark matter. For comparison, consider the dwarf spheroidal galaxy Draco. Assuming a smooth NFW halo, dark matter particles could be annihilating within Draco at a rate as large as $\sim 5 \times 10^{35} \, {\rm s}^{-1} \, (600 \, {\rm GeV}/m_{\chi})^2 \, (\sigma v/3\times 10^{-26} {\rm cm}^3/{\rm s})$~\cite{draco}. Thus, even with its very substantial mass of approximately $3 \times 10^7$ solar masses, a Draco-like object located within a fraction of a kpc from the Solar System would be only marginally capable of generating the PAMELA and ATIC signals for a thermal WIMP distributed throughout the clump with a smooth NFW profile. 

The mass and density required of the clump could be relaxed considerably, however, if the neutralino were produced non-thermally in the early universe.  In particular, an 600 GeV wino-like neutralino is predicted to have an annihilation cross section of $\sigma v \approx 3 \times 10^{-25}$ cm$^3$/s, which is approximately an order of magnitude larger than the value predicted for a typical thermal relic. Alternatively, inhomogeneities within the clump's dark matter distribution can lead to a higher annihilation rate.  If some combination of these effects enhance the annihilation rate by a factor of $\sim$10, then the calculations of Ref.~\cite{taylor} can be used to estimate that there is a 0.2\% to 0.02\% chance of a sufficiently massive and/or dense clump being close enough to the Solar System to generate the observed cosmic ray electrons and positrons.

Lastly, we turn our attention to the constraints that can be placed from observations of gamma rays, synchrotron emission, and cosmic ray antiprotons. For a typical distribution of dark matter throughout the halo of the Milky Way, the strongest indirect detection constraints come from limits on the gamma ray~\cite{dingus} and synchrotron~\cite{syn} flux from the inner galaxy. In the case in which the PAMELA and ATIC signals are dominated from nearby annihilations, the annihilation rate in the inner Milky Way can easily be low enough to evade these constraints. A nearby clump of neutralinos would, however, produce an extended source of gamma rays which could potentially have been observed by EGRET (or in the future by FERMI/GLAST). Considering a clump of 800 GeV neutralinos annihilating at a rate of $1.7\times 10^{37}$ s$^{-1}$, 1 kpc from the Solar System, we estimate the flux of gamma rays at 1 GeV to be $E_{\gamma}^2 dN_{\gamma}/dE_{\gamma} \approx 5 \times 10^{-6} \, {\rm GeV}\, {\rm cm}^{-2}\, {\rm s}^{-1}\, {\rm sr}^{-1} \, (200\, {\rm pc}/r)^2$, and at 10 GeV to be $E_{\gamma}^2 dN_{\gamma}/dE_{\gamma} \approx 2 \times 10^{-5} \, {\rm GeV}\, {\rm cm}^{-2}\, {\rm s}^{-1}\, {\rm sr}^{-1} \, (200\, {\rm pc}/r)^2$, where $r$ is the radial size of the annihilating region. As this is below or comparable to the diffuse background measured by EGRET~\cite{egretbg}, the presence of such a clump appears to be consistent with this constraint.

In addition to measuring the positron fraction, PAMELA has also published their measurement of the antiproton-to-proton ratio in the cosmic ray spectrum~\cite{pamelaantiproton}. This result, which is consistent with pure secondary production of antiprotons, can be used to constrain the rate of dark matter annihilations in the Milky Way halo. Ref.~\cite{antiprotons2}, for example, found that if TeV-mass WIMPs annihilating to $W^+ W^-$ throughout the halo of the Milky Way is normalized to produce the PAMELA positron fraction, then the flux of antiprotons would exceed the observed antiproton-to-proton ratio by about an order of magnitude. This conclusion is altered considerably in the case of a nearby annihilating clump, however.  600 GeV (100 GeV) electrons and positrons typically lose most of their energy by the time they have traveled $\sim 1$ kpc (a few kpc). In contrast, antiprotons propagate without significant energy losses, and can contribute to the cosmic ray spectrum even if they originate from very distant regions of the galaxy. As a result, the constraint from cosmic ray antiprotons is relaxed considerably when only the local annihilation rate is normalized to the PAMELA positron fraction. Furthermore, one should keep in mind that modifications to the propagation model (especially the width of the diffusion zone) could lead to an antiproton constraint which is less stringent by a factor of 20 or more than described in Ref.~\cite{antiprotons2}.

In summary, we have discussed in this letter the possibility that the excesses recently observed by PAMELA and ATIC may be the product of a nearby clump of annihilating neutralino dark matter. This scenario solves three problems that are typically faced when attempting to explain these signals with annihilating neutralinos. First, the spectrum is hardened considerably (especially when the motion of the clump is properly accounted for), bringing the predicted spectrum into line with the measurements of PAMELA and ATIC.  Second, a larger annihilation rate is expected, making the observed flux less challenging to accomodate.  Third, the constraints from cosmic ray antiprotons are relaxed considerably, as well as those from gamma rays and synchrotron emission from the inner Milky Way.

To determine whether or not a nearby clump of annihilating neutralinos is in fact responsible for the ATIC and PAMELA signals, further data will be required. In addition to more data from PAMELA (at higher energies, and with greater exposure), ground based gamma ray telescopes such as HESS and VERITAS should be capable of measuring the cosmic ray electron spectrum over the energy range of the ATIC feature with higher precision than is currently available~\cite{act}. 

Data from the FERMI/GLAST gamma ray telescope will be valuable in testing this hypothesis in two ways. Firstly, this experiment may be capable of detecting gamma rays from the clump itself, which would provide a smoking gun for the scenario described in this letter.  Secondly, FERMI will also detect very large nubmers of cosmic ray electrons/positrons, which could potentially enable the detection of a small ($\sim$0.1\%) dipole anisotropy in their angular distribution~\cite{pulsars}.  Such an anisotropy would not be expected if dark matter annihilating throughout the halo were responsible for the PAMELA and ATIC signals.

\medskip

This work has been supported by the US Department of Energy, including grant DE-FG02-95ER40896, and by NASA grant NAG5-10842.

\end{document}